\documentclass[]{aa}
\usepackage{epsfig}
\usepackage{graphicx}
\usepackage{subfigure}
\begin{document}

\title{Photometric Mapping with ISOPHOT using the
``P32'' Astronomical Observation Template
\thanks{Based on observations with {\em ISO}, an {\em ESA} project with instruments funded by {\em ESA} 
member states (especially the P/I countries France, Germany, the Netherlands and the United Kingdom)
with participation of {\em ISAS} and {\em NASA}.}
}
\authorrunning{Tuffs \& Gabriel 2003}
\titlerunning{Photometric Mapping with ISOPHOT}

\author{R.\,J.\,Tuffs\inst{1} \and C.\,Gabriel\inst{2}}
  \institute{Astrophysics Division, Max-Planck-Institut f\"ur Kernphysik,
     Saupfercheckweg 1, 69117 Heidelberg, Germany. 
     Richard.Tuffs@mpi-hd.mpg.de
  \and
     ISO Data Centre, Astrophysics Division,
     ESA, Villafranca del Castillo, P.O. Box 50727, 28080 Madrid, Spain.
     cgabriel@xvsoc01.vilspa.esa.es
     \thanks{now at XMM-Newton Science Operations Centre,
ESA, Villafranca del Castillo, P.O. Box 50727, 28080 Madrid, Spain}
}

\offprints{Richard.Tuffs@mpi-hd.mpg.de}

\date{Received; accepted}

\abstract{
The ``P32'' Astronomical Observation Template (AOT) provided a means to
map large areas of sky (up to 45$\times$45 arcmin) in the 
far-infrared (FIR) at 
high redundancy and with sampling close to the Nyquist limit 
using the ISOPHOT C100 (3$\times$3) and C200 (2$\times$2)
detector arrays on board the Infrared Space Observatory (ISO). 
However, the transient response behaviour of the Ga:Ge detectors,
if uncorrected, can lead to severe systematic photometric errors and 
distortions of source morphology on maps.
We describe the basic concepts of an algorithm which can
successfully correct for transient response artifacts in P32 observations. 
Examples are given to demonstrate the photometric and imaging performance 
of ISOPHOT P32 observations of point and extended sources corrected using the
algorithm. For extended sources we give the integrated flux densities of
the nearby galaxies NGC~6946, M~51 and M~101. and an image of M~101
at 100$\,\rm \mu m$.


}

\maketitle

\section{Introduction}
From the point of view of signal processing and photometry,
diffraction-limited mapping in the FIR with cryogenic space 
observatories equipped with photoconductor 
detectors poses a particular challenge. In this wavelength regime 
the number of pixels in detector arrays is limited in comparison with 
that in mid- and near-IR detectors. This means that more repointings 
are needed to map structures spanning a given number of 
resolution elements. Due to the logistical constraints imposed
by the limited operational lifetime of a cryogenic mission, this
inevitably leads to the problem that the time scale for modulation of 
illumination on the detector pixels becomes smaller than 
the characteristic transient response timescale of the detectors to 
steps in illumination. The latter timescale can reach minutes
for the low levels of illumination encountered on board a space 
observatory. The result is that the signals from the detectors 
can depend as much on the illumination history as on the instantaneous 
illumination. Unless corrected for, the transient response behaviour 
of the detectors lead to distortions in images, 
as well as to systematic errors in the photometry, especially
for discrete sources appearing on the maps. In general, these 
artifacts become more long lived and more difficult to correct 
for at fainter levels of illumination, since the transient 
response timescales increase with decreasing illumination. 

For mapping observations in the FIR,
the Infrared Space Observatory (ISO; Kessler et al. 1996) was equipped 
with two Gallium-doped Germanium photoconductor detectors which 
comprised part of the ISOPHOT instrument (Lemke et al. 1996).
The ISOPHOT C200 detector was a 2$\,\times\,$2 (stressed) pixel 
array operating in the 100$\,-\,$200$\,\rm \mu m$ range, and the 
ISOPHOT C100 detector was a 3$\,\times\,$3 pixel 
array operating in the 50$\,-\,$100$\,\rm \mu m$ range.
Compared to the IRAS survey detectors, the ISOPHOT-C detectors
had relatively small pixels designed to 
provide near diffraction limiting imaging. ISOPHOT thus generally
encountered larger contrasts in illumination between source and 
background than IRAS did, making the artifacts from the transient 
response more prominent, particularly for fields with bright compact
sources and faint backgrounds.

A further difficulty specific to mapping in the FIR with ISO was
that, unlike IRAS, the satellite had no possibility to cover a target 
field in a controlled raster slew mode. This limited the field size 
that could be mapped using the spacecraft raster pointing mode alone, 
since the minimum time interval between the satellite fine pointings used
in this mode was around 8\,s. This often greatly exceeded the nominal 
exposure time needed to reach a required level of sensitivity 
(or even for many fields the confusion limit).
Furthermore, the angular sampling and redundancy achievable using the
fine pointing mode in the available time was often quite limited, 
so that compromises sometimes had to be made to adequately extend 
the map onto the background.

A specific operational mode for ISO - the ``P32'' Astronomical 
Observation Template (AOT) - was developed 
for the ISOPHOT instrument to alleviate 
these effects (Heinrichsen et al. 1997). 
This mode employed a combination of standard 
spacecraft repointings and rapid oversampled scans 
using the focal plane chopper. The technique could 
achieve a Nyquist sampling on map areas of sky ranging up to 
45$\times$45 
arcmin in extent (ca. 70$\times$70 FWHM resolution elements) on timescales 
of no more than a few hours. In addition to mapping large sources, the 
P32 AOT was extensively used to observe very faint compact 
sources where the improved sky sampling and redundancy alleviated 
the effects of confusion and glitching.

In all, over 6$\%$ of the observing time of ISO
was devoted to P32 observations during the 1995-1998 mission, 
but the mode could not until now be fully exploited scientifically
due to the lack of a means of correcting for the complex non-linear 
response behaviour of the Ge:Ga detectors. Here we
describe the basic concept of an algorithm which can
successfully correct for the transient response artifacts in 
P32 observations. This algorithm forms the kernel of the ``P32Tools'' 
package, which is now publically available as
part of the ISOPHOT Interactive Analysis package PIA (Gabriel et. al 1997;
Gabriel \& Acosta-Pulido 1999).
Information on the algorithm, as well as the first scientific applications, 
can also be found in Tuffs et al. (2002a,b) and Tuffs \& Gabriel (2002).
The user interface of P32Tools that connects the P32 algorithm to PIA
is described by Lu et al. (2002).
After a brief overview of relevant aspects of the P32 AOT in Sect.~2,
we describe the semi-empirical model used to reproduce the transient 
response behaviour of the PHT-C detectors in Sect.~3. Sect.~4 
describes the algorithms used to correct data. Examples demonstrating
the photometric and imaging performance of ISOPHOT P32 observations
are given in Sect.~5, based on maps corrected using the P32 algorithm.

\section{The P32 AOT}

The basic concept of the P32 AOT, in which imaging was
achieved using a combination of standard spacecraft repointings 
and rapid oversampled scans using ISOPHOT's focal plane chopper, 
is summarised in Fig.~1.
\begin{figure}[htb]
\includegraphics[scale=0.45,angle=0.]{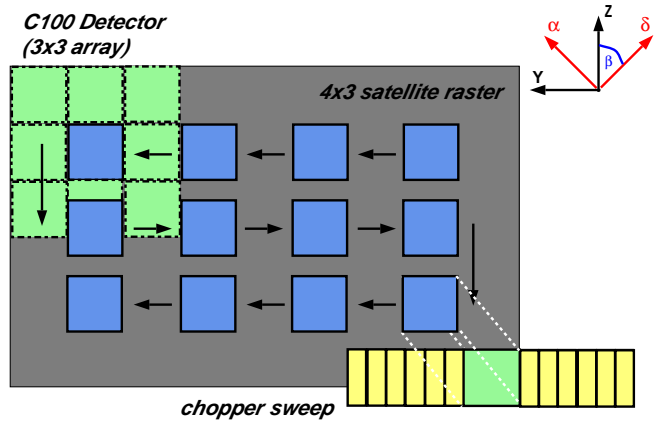}
    \caption{A schematic representation of the P32 observing technique.
The sky is sampled using a combination of a coarse spacecraft raster
in the satellite Y and Z coordinates (horizontal and vertical
directions on the figure) and finely sampled sweeps of the focal
plane chopper in the Y coordinate.
The spacecraft raster (in this example a 4$\,\times\,$3 raster) 
starts at the maximum value of Z and the minimum
value of Y and proceeds as shown by the arrows. The position angle
$\beta$ of the first scan leg in Y according to normal astronomical convention
(taken positive E from N) is related to the satellite roll angle $\alpha$ by
$\beta$~=~$\alpha$~+~$90^{o}$.
At each spacecraft pointing direction
a chopper sweep of length 180 arcsec (the diameter of the unvignetted
field of view of ISO) is made with a sampling of 
1/3 of the detector pixel separation. This results in
chopper sweeps of 13 positions separated by 15 arcsec for 
observations using the C100 detector array (as depicted in this example), 
and chopper sweeps of 7 positions sampled at 31 arcsec for the 
C200 detector array. The spacecraft raster sampling interval in
Y is set to be a multiple of the chopper sampling interval
(in the example depicted it is set to 6 chopper sampling
intervals). A series of uniformly sampled
linear scans for each detector pixel in Y is obtained in this fashion, 
built up from a sequence of registered and overlapping chopper sweeps 
obtained on each spacecraft scan leg in Y. The sampling of the target 
field in Z is controlled entirely by the spacecraft raster sampling
interval in Z. In the example depicted the C100 detector
detector array is stepped in intervals of 1.5 times the
detector pixel separation in Z. For this particular case,
the linear scans from different C100 detector pixels can be
combined to sample the sky at half the detector pixel separation in Z,
resulting in an overall sampling of 15$\times$23 arcsec over the 
target field.}
\end{figure}
There were two basic requirements, outlined in the original proposal 
for the AOT (Tuffs \& Chini 1990) which led to the final design:
The first goal was to give ISOPHOT the capability of
achieving a Nyquist sky sampling 
($\Delta\Theta\,$=$\,\lambda$/2D$\,$=$\,$17$\,$arcsec at 100$\,\rm \mu m$)
on extended sources subtending up to 50 resolution elements
in a tractable observation time. As well as providing a unique 
representation of any arbitrary sky brightnesss distribution, this 
also improved the effective confusion limit in deep observations, 
which, in directions of low background was $\sim100$ 
and 20\,mJy rms at 200 and 100$\,\rm \mu m$, respectively, 
for ISOPHOT-C. 

The second requirement
was to achieve a redundancy in the observed data. This led to the
sampling of each of a given set of sky directions by each detector pixel 
at different times. The redundancy was achieved in part
by making several (at least 4) chopper sweeps at each spacecraft
pointing direction. Further redundancy was achieved by having 
an overlap in sky coverage of at least 
half the chopper sweep amplitude between chopper sweeps made at successive 
spacecraft pointing directions in the Y spacecraft coordinate. Thus,
a given sky direction was sampled on three different time scales
by each detector pixel: on intervals of the detector non-destructive
read interval, on intervals of the chopper sweep, and on intervals
of the spacecraft raster pointings. Although the redundancy
requirement was originally made to help follow the long term trends 
in detector responsivity during an observation, in practice it also 
proved very useful in removing effects caused by cosmic particle
impacts (Gabriel \& Acosta-Pulido 2000), 
the so-called ``glitches'', from the data. This was especially 
important to optimise the sensitivity of deep observations using the 
C100 detector, which are generally limited by glitching rather than 
by confusion.

Both these requirements led to dwell times per chopper plateau
which could be as low as 0.1\,s. For a given sky brightness, this also
had the consequence that the detector was generally read faster than in the 
other AOTs using the PHT-C detectors, which allows a higher
time resolution in the deglitching procedures.
A further beneficial side effect of the P32 AOT was that
because the transient response timescale of the detectors is
longer than the dwell time of the chopper, even for bright sources,
the effective limiting source brightness that could be 
observed without saturating the detector was higher than for
the other AOTs.

\subsection{The P32 ``natural grid''}

The fact that the spacecraft pointing increment in Y was
constrained to be a multiple of the chopper step interval
means 
that a rectangular grid of directions on the sky can be
defined such that all data taken during P32
measurement fine pointings will, to within 
the precision of the fine pointings,
exactly fall onto the directions of the grid. This so-called
P32 ``natural grid'' allows maps to be made without any gridding 
function, thus optimising angular resolution. The P32 natural
grid is also a basic building block of the algorithm to correct for the 
transient response behaviour of the detector, as the algorithm solves
for the intrinsic sky brightness in M$\times$N independent variables
for a grid dimension of M$\times$N. Here M and N are the number of
columns and rows on the P32 natural grid. An example of the
almost perfect registration of data onto the grid is shown in
Fig.~2.

All the example maps shown in this paper will be sampled without
a gridding function on the P32 natural grid. This means that the
data in each of the map pixels is independent. This is unlike 
the situation for most other maps calculated using PIA, where a
gridding function is employed and so raw data can contribute to 
more than one map pixel.

\begin{figure}[htb]
\includegraphics[scale=0.55,angle=90.]{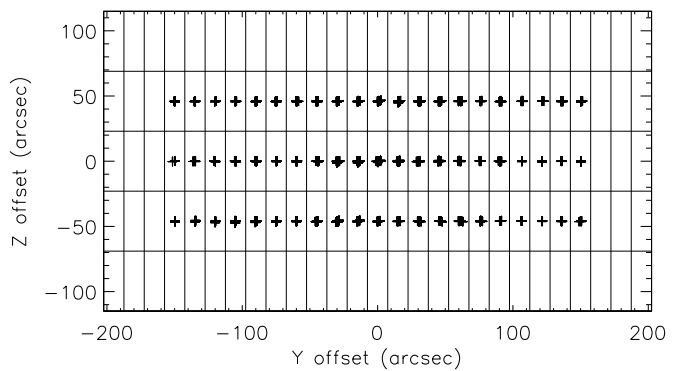}
\caption{Pointing directions observed towards Ceres
at 105$\,{\mu}m$. The spacecraft raster dimension was $3\,\times\,3$ in 
the spacecraft coordinates $Y\,\times\,Z$. Sky sampling in $Y$ is
achieved through a combination of chopper scans in $Y$ and 
repointings of the spacecraft. In $Z$ the sky sampling is determined
alone by the spacecraft repointing interval. Each pointing direction 
for 11833 individual data samples for the central pixel of the C100 
detector array is plotted as a cross. The rectangular grid
is the ``P32 natural grid'' (see text) for this observation, on which
the sky brightness distribution is to be solved. The grid sampling
is $14.99\times45.98$ arcsec. Only points with the on target flag set
(i.e. not including slews) are plotted.
The pointings typically lie within 1 arcsec of the centre of 
each pixel of the P32 natural
grid.}
\end{figure}

\section{Transient response of the PHT-C detectors}

Gallium doped germanium photoconductor detectors exhibit a number of effects
under low background conditions which create severe calibration
uncertainties. The ISOPHOT C200 (Ge:Ga stressed) and especially the C100
\begin{figure}[htb]
\includegraphics[scale=0.52]{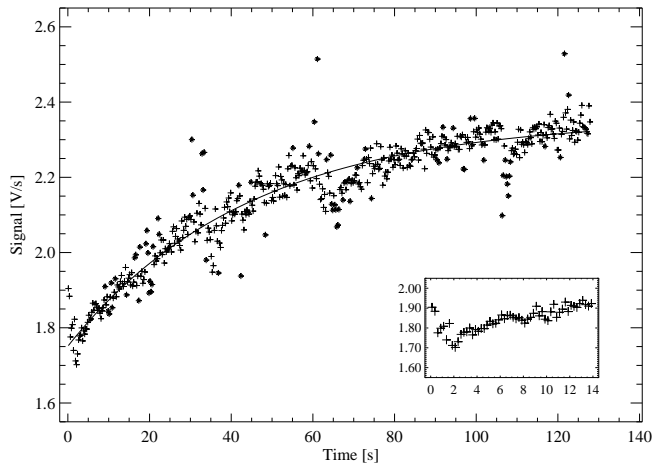}
\caption{Example of the transient response of a Ge:Ga detector
following an upwards step to a constant level of illumination
at 100$\,\rm \mu m$. The hook response is shown in the inset.}
\end{figure}
(Ge:Ga) detectors have a complex non-linear response
as a function of illumination history on timescales of $\sim0.1-100$\,sec, 
which depends on the absolute illumination as well as
the changes in illumination. The following behaviour is seen 
in response to illumination steps (see also Fouks 1992, Acosta-Pulido et al. 
2000 and Laureijs et al. 2000, Sect 4.2.2.):

\begin{itemize}

\item a rapid jump in signal, which however undershoots
the asymptotic equilibrium response (often severely for the C100 array)

\item a hook response (overshooting or undershooting on intermediate
timescales of $\sim1-10$\,s)

\item a slow convergence to the asymptotic equilibrium response of 
the detector on timescales of $\sim10-100$\,s

\end{itemize}
An example showing these effects for a staring observation
is given in Fig.~3. Also seen are points discarded by the deglitching
algorithm in PIA, marked by stars. Many glitches are 
accompanied by a longer lived ``tail''.
It is these tails that determine the
ultimate sensitivity of the ISOPHOT C100 array.

\subsection{Semi-empirical detector model} 

Unlike the ISOCAM and ISOPHOT-S Si:Ge detectors on board ISO, which 
operate in the mid-IR, the transient response behaviour of the 
Ge:Ga detectors of ISOPHOT-C are not described by the analytical 
Fouks Schubert model (Fouks \& Schubert 1995; Schubert et al. 1995; 
Acosta-Pulido 1998; see Coulais \& Abergel 2000 for its application to
ISOCAM). Although numerical simulations of
the transient behaviour of Ge:Ga detectors have been made
(Hagel et al. 1996), these do not cover the wide range of 
operating conditions experienced by the ISOPHOT detectors. 
However, the effects observed in ISOPHOT-C
can to some extent be qualitatively understood 
in terms of the build up of charge on the contacts between 
semiconductor and readout circuit, as analytically modelled by Sclar (1984).
Sclar's model can be formulated as the superposition of three 
exponentials to describe the response of the detector to a step in
illumination. A superposition of three exponentials has also
been proposed by Fujiwara et al. (1995) and Church et al. (1996) 
to represent the transient behaviour. The three timescales involved 
are a minimum requirement to reproduce the hook response and the 
long term approach to the asymptotic response.  
However, no theory exists to predict values for the constants in this 
representation, and thus they must be found empirically. 

In practice it is a formidable and unsolved problem to extract the constants 
from in flight data for as many as three temporal components to the
transient response behaviour. 
In this paper we describe 
an approximate solution involving a superposition of two exponentials, 
each with illumination-dependent constants. 
This was found to adequately represent 
the detector response on timescales greater than a few seconds, though it
only approximately model the hook response (see also Blomme \& Runacres
2002). Our model can be viewed as a 
generalisation of the single exponential model developed for correction
of longer stares by Acosta-Pulido, Gabriel \& Casta\~neda (2000).

\begin{figure}[h]
    \epsfig{file=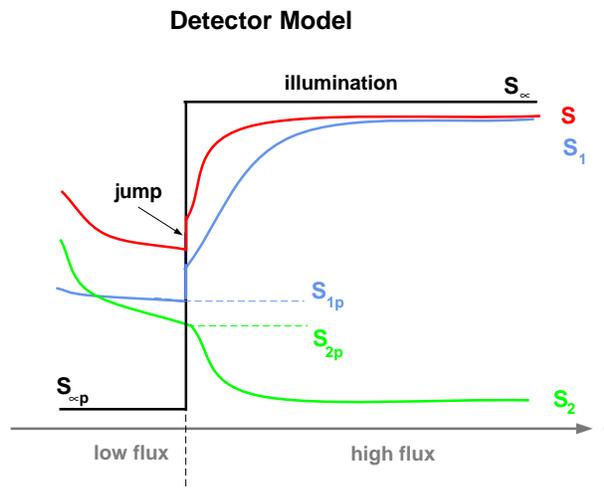, width=8cm}
    \caption{A schematic representation of the response 
     of the semi-empirical detector model used in the P32
     algorithm to an upwards step in illumination from
     $S_{\rm \infty p}$ to $S_{\rm \infty}$. The overall
     response $S$ is the sum of a ``slow'' signal response $S_{1}$ 
     and a ``fast'' signal response $S_{2}$ (see Sect.~3.1).
     $S_{\rm 1p}$ and $S_{\rm 2p}$ respectively denote the signal 
     response of these two components immediately prior to the step
     in illumination.\label{detectormodel}}
\end{figure}

In the model considered here, schematically illustrated in Fig.~4,
the basic detector response $S$
to an instantaneous step in illumination 
from $S_{\rm \infty p}$ to $S_{\rm \infty}$
is given by the sum of a ``slow'' signal response
$S_{1}$ and a ``fast'' signal response $S_{2}$ as:
\begin{eqnarray}
  S &= &S_{\rm 1} + S_{\rm 2} \\
  S_{\rm 1} &= &(1 - \beta_{\rm 2})\,S_{\rm \infty}\,
(1 - {\rm exp}[-t/\tau_{\rm 1}])\,+\,S_{\rm 01}\,{\rm exp}[-t/\tau_{\rm 1}]\\
  S_{\rm 2} &= &\beta_{\rm 2}\,S_{\rm \infty}\,
(1 - {\rm exp}[-t/\tau_{\rm 2}])\,+\,S_{\rm 02}\,{\rm exp}[-t/\tau_{\rm 2}]
\end{eqnarray}
$S_{\rm 01}$ and $S_{\rm 02}$ are signals corresponding to
the slow and fast response components immediately following 
the illumination step. $S_{\rm 01}$ and $S_{\rm 02}$ are 
related to the corresponding signals immediately prior to the
illumination step, $S_{\rm 1p}$ and $S_{\rm 2p}$, which contain
the information about the illumination history. $S_{\rm 01}$ and $S_{\rm 02}$
are specified by the jump conditions:
\begin{eqnarray}
  S_{\rm 01}&= &\beta_{\rm 1}\,(S_{\rm \infty} - S_{\rm \infty p}) + S_{\rm 1p}\\
  S_{\rm 02}&= &S_{\rm 2p}
\end{eqnarray}
The model is thus specified by four primary parameters: the slow and fast
timescales $\tau_{\rm 1}$ and $\tau_{\rm 2}$, the ``jump''
factor $\beta_{\rm 1}$ and a constant of proportionality
for the fast response $\beta_{\rm 2}$. In order to obtain satisfactory
model fits to complex illumination histories,
each of these primary parameters had to be specified as monotonic
functions of illumination, each with three subsiduary parameters:
\begin{eqnarray}
  \beta_{\rm 1} = \beta_{\rm 10}\,+\,\beta_{\rm 11}\,*\,S_{\rm \infty}^{\beta_{\rm 12}}\\
  \tau_{\rm 1} = \tau_{\rm 10}\,+\,\tau_{\rm 11}\,*\,S_{\rm \infty}^{-\tau_{\rm 12}}\\
  \beta_{\rm 2} = \beta_{\rm 20}\,+\,\beta_{\rm 21}\,*\,S_{\rm \infty}^{\beta_{\rm 22}}\\
  \tau_{\rm 2} = \tau_{\rm 20}\,+\,\tau_{\rm 21}\,*\,S_{\rm \infty}^{-\tau_{\rm 22}}
\end{eqnarray}
Thus 12 subsiduary parameters in all are needed to characterise the transient
response behaviour of the detectors. In addition, the initial starting 
state of the detector, given by the 
values of $S_{\rm 1p}$ and $S_{\rm 2p}$ immediately prior to
the start of the mapping observation, must be specified.

\begin{table*}[htb]
\caption{Model parameters for the C100 detector}
\begin{tabular}{l|ccccccccc}
\hline\hline
parameters      & \multicolumn{9}{c}{C100 detector pixels}\\
\hline
                & 1      & 2      & 3      & 4      & 5      & 6      & 7      & 8      & 9 \\
\hline
$\beta_{10}$ &  0.995 &  6.100 &  2.170 &  1.200 &  2.120 &  6.680 &  4.630 &  0.960 &  2.190\\
$\beta_{11}$ & -0.69  & -5.36  & -1.52  & -0.56  & -1.82  & -5.96  & -3.95  & -0.28  & -1.89\\
$\beta_{12}$ &  0.059 &  0.023 &  0.049 &  0.092 &  0.022 &  0.018 &  0.032 &  0.075 &  0.036\\
$\tau_{10}$  &  6.16  &  5.80  &  7.50  &  6.63  &  6.92  &  5.07  &  5.72  &  7.73  &  8.60\\
$\tau_{11}$  &  7.75  & 17.25  & 12.90  & 12.41  &  4.28  & 12.34  & 12.69  & 11.60  &  1.04\\
$\tau_{12}$  & -0.65  & -1.28  & -1.04  & -0.88  & -1.22  & -0.65  & -0.88  & -1.28  & -2.32\\ 
$\beta_{20}$ &  0.661 &  5.866 &  5.868 &  0.732 & -0.534 &  6.490 &  4.400 &  1.171 &  0.140\\
$\beta_{21}$ & -0.488 & -5.520 & -5.515 & -0.423 &  0.723 & -6.11  & -4.133 & -0.870 &  0.000\\
$\beta_{22}$ &  0.02840 & 0.00814 & 0.00434 & 0.03950 & -0.01030 & 0.00459 & 0.01140 & -0.01450 & 0.00000\\
$\tau_{20}$  &  0.376 & 0.301  & 0.388  & 0.330  & 14.890 & 0.766  & 0.664  & 0.333  & 0.605\\
$\tau_{21}$  &  0.324 & 0.257  & 0.305  & 0.368  & -14.240 & 0.647 & 0.139  & 0.381  & 0.577\\
$\tau_{22}$  &  0.38400 & 0.53700 & 0.60300 & 0.60500 & 0.01025 & 0.55100 & 0.65200 & 0.58400 & 0.43900\\
\hline
\end{tabular}
\end{table*}
\vspace{0.7cm}

\begin{table}
\caption{Model parameters for the C200 detector}
\begin{tabular}{l|cccc}
\hline\hline
parameters     & \multicolumn{4}{c}{C200 detector pixels}\\
\hline
              & 1     & 2     & 3     & 4\\
\hline
$\beta_{10}$  &  0.94 &  0.98 &  0.86 &  1.01\\
$\beta_{11}$  & -0.12 & -0.16 & -0.10 & -0.14\\
$\beta_{12}$  &  0.23 &  0.20 &  0.22 &  0.27\\
$\tau_{10}$   &  5.92 &  4.53 &  3.77 &  4.92\\
$\tau_{11}$   &  4.65 &  6.68 &  5.34 &  5.46\\
$\tau_{12}$   & -0.60 & -0.49 & -0.52 & -0.57\\
$\beta_{20}$  & -0.2980 & -0.0879 & -0.1430 & -0.0269\\
$\beta_{21}$  & 0.440 & 0.245 & 0.342 & 0.200\\
$\beta_{22}$  & 0.0088 & -0.1900 & -0.0750 & -0.0241\\
$\tau_{20}$   & -4.90 & -4.87 & -4.88 & -4.95\\
$\tau_{21}$   & 5.14 & 5.20 & 5.20 & 5.14\\
$\tau_{22}$   & -0.00313 & -0.00439 & -0.00167 & -0.00249\\
\hline
\end{tabular}
\end{table}

\begin{figure}[htb]
\includegraphics[scale=0.52]{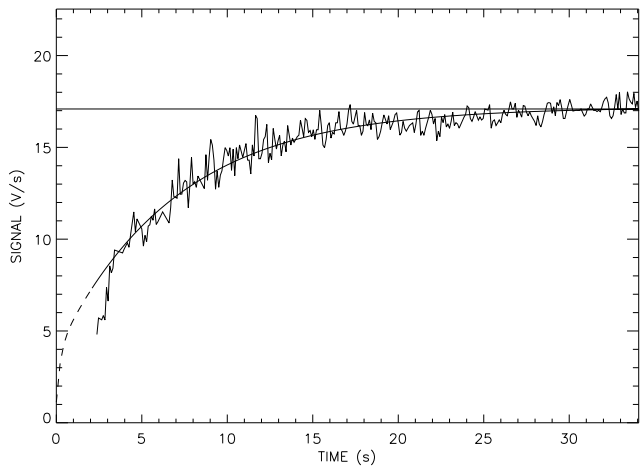}
\caption{Example of a model fit (curve) to the response
of the central pixel of the C100 array to a constant FCS illumination 
starting at $t=0$\,s. Prior to the FCS illumination the detector 
was viewing blank sky. The horizontal line indicates the fitted
value of the illumination. No data was recorded in the first two
seconds following FCS switch-on.
}
\end{figure}

\subsection{Determination of model parameters}

The parameterisation of the 
illumination dependence of the primary parameters
$\tau_{\rm 1}$, $\beta_{\rm 1}$,
$\tau_{\rm 2}$ and $\beta_{\rm 2}$ was made in 
engineering units (V/s), rather than in astrophysical units such  
as MJy/sr. Thus, it was assumed that there is no dependence of 
the detector response on the wavelength of the IR photons reaching 
the detector. Since the individual pixels of the C100 and the 
C200 arrays behave as independent detectors, each of the 12 
parameters of the detector model had to be separately determined 
for each detector pixel.
 
The ``slow'' primary parameters $\tau_{\rm 1}$ and $\beta_{\rm 1}$ were
found from fits to the detector response to fine calibration
source (FCS) exposures. To make this determination independent
of the determination of the ``fast'' parameters, only data points 
corresponding to times of 5s or longer after the switch-on of the FCS 
was used. An example of a fit to the signal from an FCS exposure
is given in Fig.~5. The model fit
to the signal response on time scales ranging from a few seconds up to the 
duration of the FCS exposure depends mainly on the values of 
$\tau_{\rm 1}$ and $\beta_{\rm 1}$. 

A database of values for
$\tau_{\rm 1}$ and $\beta_{\rm 1}$ was built up for a sample
of FCS measurements, selected so as to correspond to the complete
range of sky and source illuminations encountered by ISOPHOT during
the ISO mission. These allowed the dependence 
of $\tau_{\rm 1}$ and $\beta_{\rm 1}$ on illumination 
to be determined according to Eqns.
6$\,\&\,$7. An example showing the illumination dependence of $\beta_{\rm 1}$
for the central pixel of the C100 array is given in Fig.~6a. The
solid line is the fitted illumination dependence of $\beta_{\rm 1}$
used to find the values of the parameters $\beta_{\rm 10}$,
$\beta_{\rm 11}$ and $\beta_{\rm 12}$ used in the model. A corresponding
plot for the illumination dependence of $\tau_{\rm 1}$ is given in
Fig.~6b.

\begin{figure*}
\subfigure[]{
\includegraphics[scale=0.74]{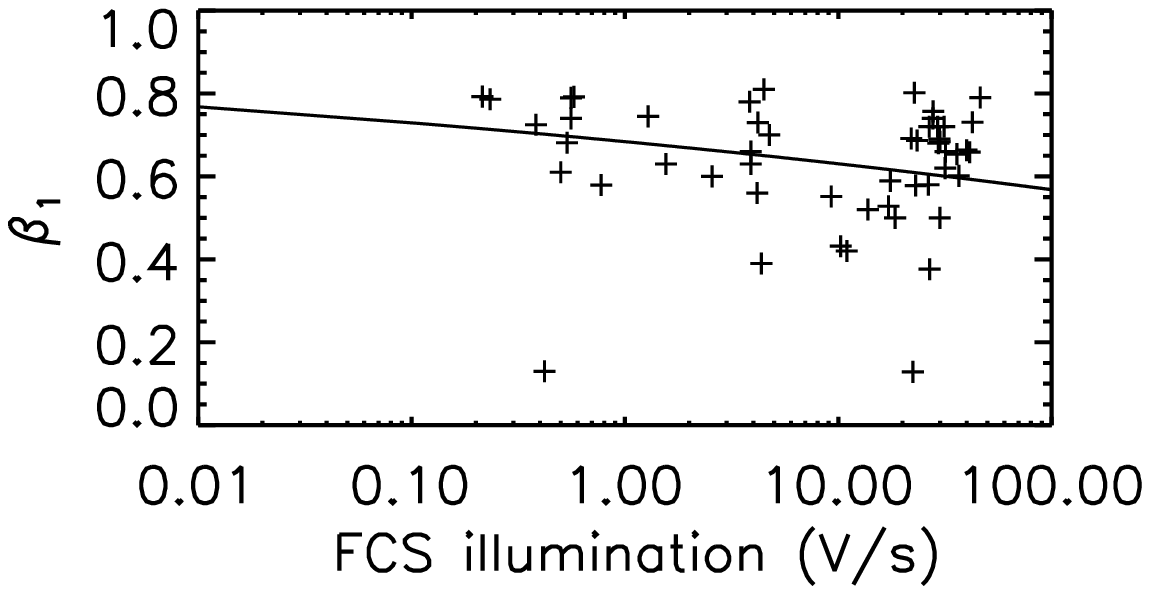}
}
\subfigure[]{
\includegraphics[scale=0.74]{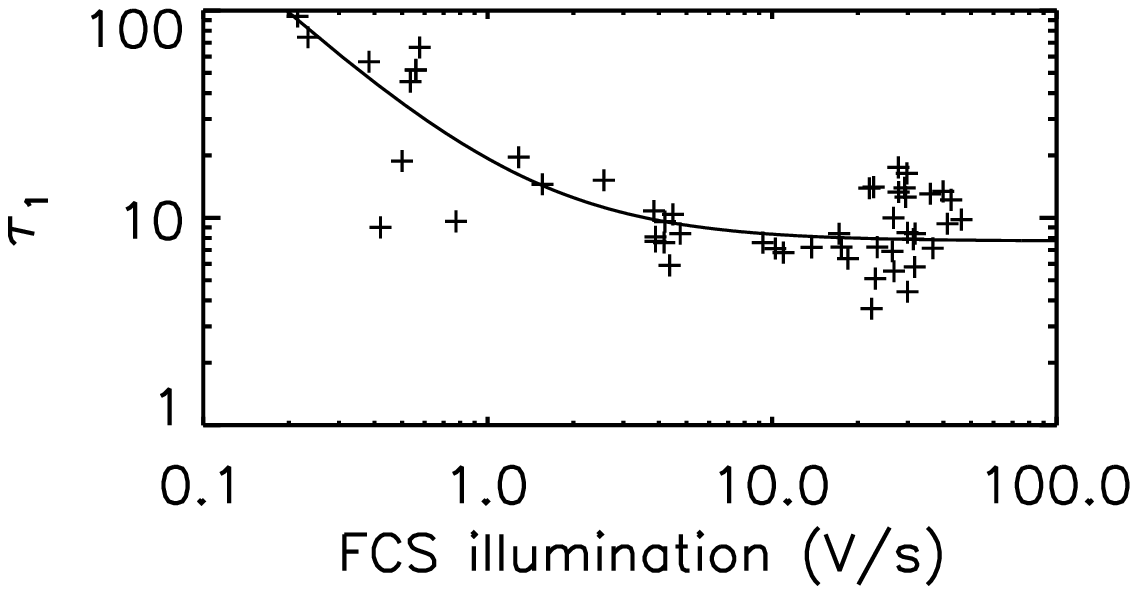}
}
\subfigure[]{
\includegraphics[scale=0.74]{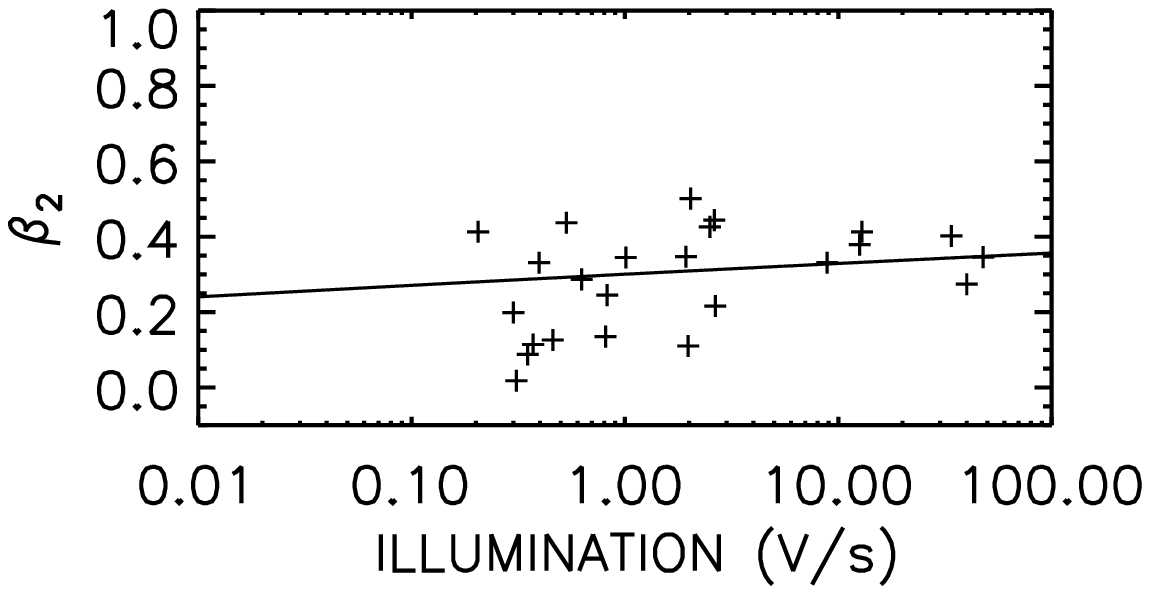}
}
\subfigure[]{
\includegraphics[scale=0.74]{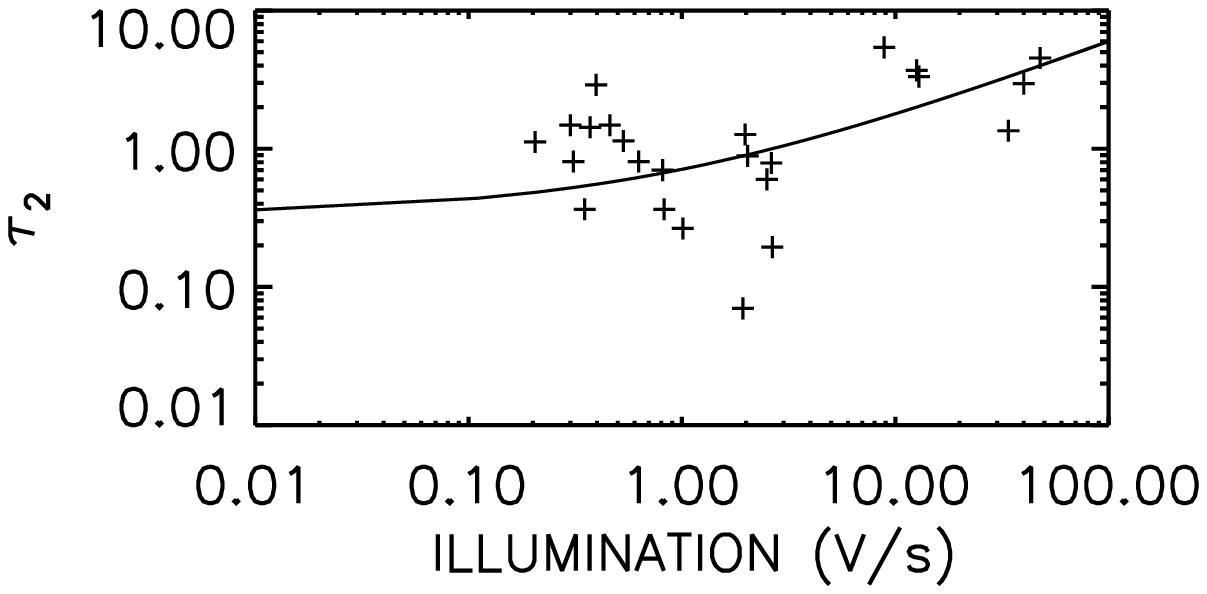}
}
\caption{a) Values of the parameter $\beta_{\rm 1}$ for pixel~8
of the C100 array found from fits to the detector response 
to FCS exposures, plotted versus FCS illumination. 
The solid line is a fit of the function 
$\beta_{\rm 1}=\beta_{\rm 10}+\beta_{\rm 11}S_{\rm \infty}^{\beta_{\rm 12}}$ 
with $\beta_{\rm 10}=0.96$,  $\beta_{\rm 11}=-0.28$,
and $\beta_{\rm 12}=0.075$. These are the parameters adopted for the
model giving the default illumination dependence of $\beta_{\rm 1}$
for this pixel.
b) As a), but for the parameter $\tau_{\rm 1}$ (in seconds).
The solid line is a fit of the function 
$\tau_{\rm 1}=\tau_{\rm 10}+\tau_{\rm 11}S_{\rm \infty}^{-\tau_{\rm 12}}$ 
with $\tau_{\rm 10}=7.73s$,  $\tau_{\rm 11}=11.60$,
and $\tau_{\rm 12}=-1.28$. These are the parameters adopted for the
model giving the default illumination dependence of $\tau_{\rm 1}$
for this pixel.
c) Values of the parameter $\beta_{\rm 2}$ for the central pixel
of the C100 array found from self calibration optimisation of
the model on observations of standard point source calibrators,
plotted versus illumination.  
The solid line is a fit of the function 
$\beta_{\rm 2}=\beta_{\rm 20}+\beta_{\rm 21}S_{\rm illum}^{\beta_{\rm 22}}$ 
with $\beta_{\rm 20}=1.171$,  $\beta_{\rm 21}=-0.870$,
and $\beta_{\rm 22}=-0.0145$, where $S_{\rm illum}$ is the known
illumination of the detector pixel due to the point source calibrator.
These are the parameters adopted for the
model giving the default illumination dependence of $\beta_{\rm 2}$
for this pixel.
d) As c), but for the parameter $\tau_{\rm 2}$ (in seconds).  
The solid line is a fit of the function 
$\tau_{\rm 2}=\tau_{\rm 20}+\tau_{\rm 21}S_{\rm illum}^{-\tau_{\rm 22}}$
with $\tau_{\rm 20}=0.333\,s$,  $\tau_{\rm 21}=0.381$,
and $\tau_{\rm 22}=0.584$.
These are the parameters adopted for the
model giving the default illumination dependence of $\tau_{\rm 2}$
for this pixel.
}
\end{figure*}

At shorter timescales the response is primarily determined by the 
``fast'' primary parameters $\tau_{\rm 2}$ and $\beta_{\rm 2}$. 
However, the FCS exposures could not be used to find $\tau_{\rm 2}$ and 
$\beta_{\rm 2}$ since no data was recorded in the first 
two seconds following 
the detector switch-on. This is also the reason why the hook response
is not seen in Fig.~5. Furthermore, there is very probably 
a non-negligible timescale
needed by the FCS to reach its final operating temperature.
Therefore, the ``fast'' parameters (with timescales 
$\sim\,$0.1$\,\le\,\tau\,\le\,\sim\,$5\,s)
were found using a self calibration technique (described in Sect.~4.2),
operating on co-added repeated chopper sweeps crossing standard point source 
calibrators. During the self calibration the ``slow'' parameters were fixed 
to the values determined in the fits to the FCS exposures.
Observations of standard point source calibrators
(12 source-wavelength combinations for both the C100 and C200 arrays)
were analysed using the self-calibration technique. Each 
observation of a source provided several points for the
determination of the "fast" detector constants, since individual 
subscans through the source could be processed independently. In all,
self calibrations were performed on data spanning a range 
in detector illumination of about 100. 
This allowed the dependence of $\tau_{\rm 2}$ and 
$\beta_{\rm 2}$ on illumination to be determined according to the 
parameterisation of Eqns. 8$\,\&\,$9.
Examples showing the illumination dependence of $\beta_{\rm 2}$
and $\tau_{\rm 2}$ for the central pixel of the C100 array are
given in Figs.~6c and 6d, respectively.
The values of the 12 parameters determined for each pixel
of the C100 and C200 detectors are given in Tables~1 and 2, respectively.

\section{The P32 algorithm}

In a standard reduction of ISOPHOT data using the interactive analysis 
procedures of PIA, the analysis is made in several irreversible steps,
starting with input ``edited raw data'' at the full time resolution,
and finishing with the final calibrated map. To correct for 
responsivity drift effects, however,
it is necessary to iterate between a sky map and the input data at full
time resolution. This different concept required
a completely new data reduction package. In its development phase, and 
for the determination of the detector model parameters, this 
data reduction package was run as a set of IDL scripts.
Running in the PIA environment, P32Tools provides access to the main
functionality of this data reduction package via a GUI interface 
(Lu et al. 2002). 

There are three basic elements to the concept of the P32 algorithm. 
The first step is signal conditioning, which aims to 
provide a stream of signal values,
each with an attached time, from which artifacts such as 
glitches, dark current, and non-linearity effects have been removed, but 
which retains the full imprint of the transient response of
the detector to the illumination history at the sharpest available
time resolution. We will refer to such a stream of signals as the
``signal timeline''.  

The second step is the transient correction itself, which is an
iterative process to determine the most likely sky brightness 
distribution giving rise to the observed signals. 
This is a non-linear optimisation problem with the values of
sky brightness as variables. For large maps several hundred independent
variables are involved.

Lastly, there is the calculation  of calibrated transient corrected
maps. This entails the conversion from V/s to MJy/sr, flat fielding,
and the coaddition of data from different detectors pixels. Using
P32Tools, this third stage requires the exporting of the transient corrected
signal timeline into PIA at the SCP level, followed by the application
of standard PIA processing. Here we just describe the new procedures
for signal conditioning and transient correction.

\begin{figure*}[htb]
\includegraphics[scale=0.8]{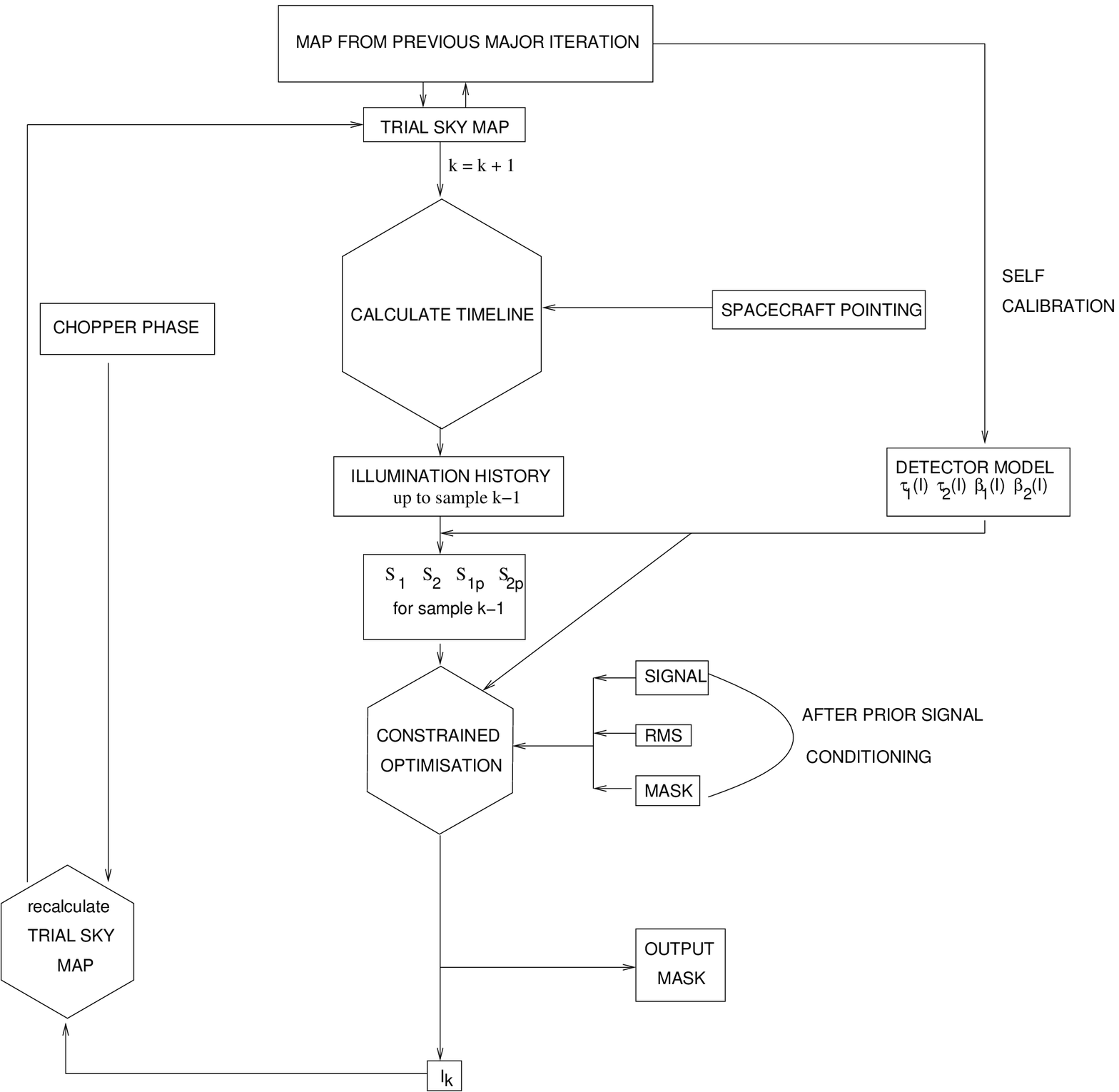}
\caption{Iterative scheme for transient correction (see Sect.~4.2)}
\end{figure*}

\subsection{Signal Conditioning}

The processing steps for signal conditioning are as follows:

\begin{figure*}[!ht]
  \begin{center}
    \epsfig{file=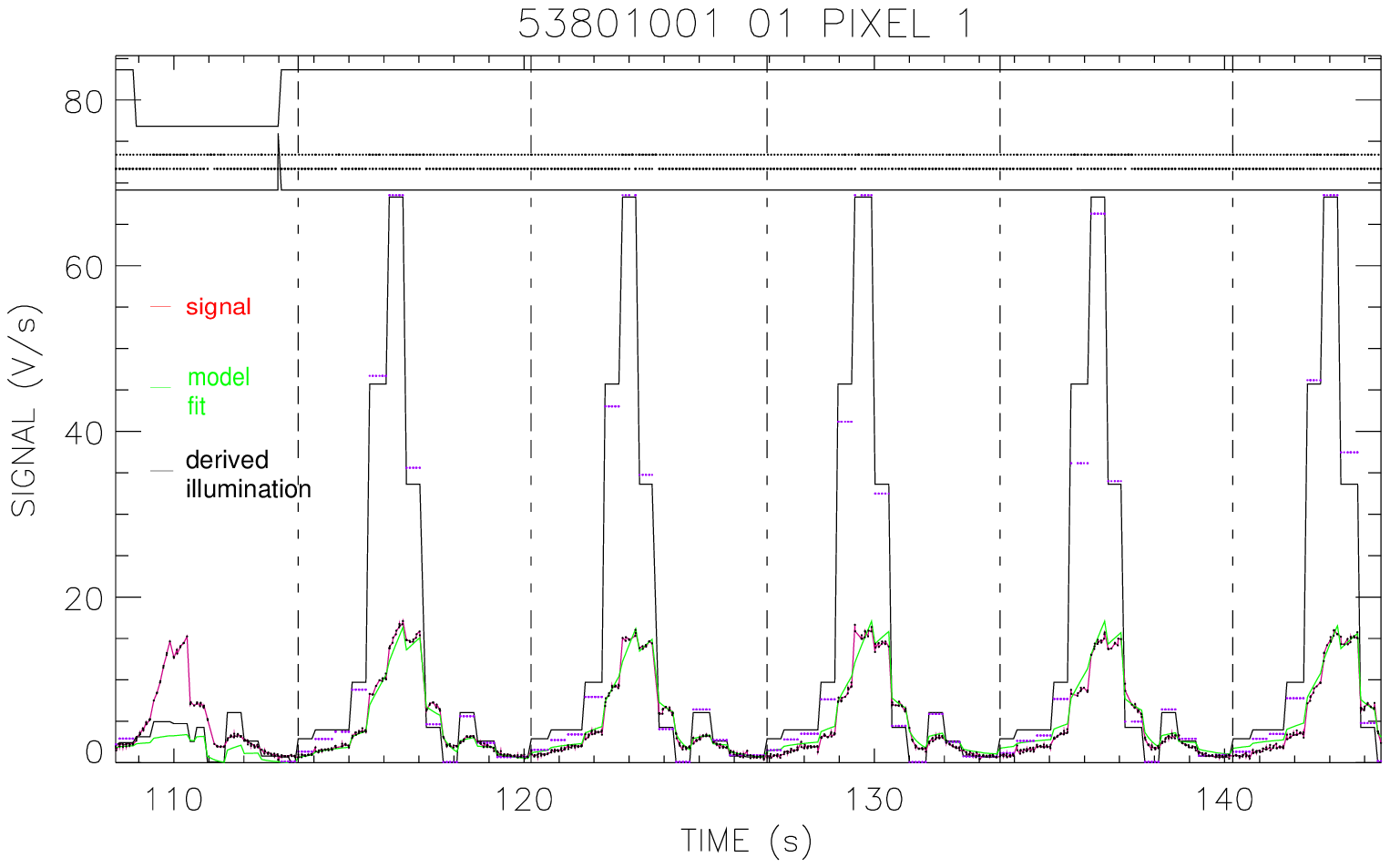, width=15.0cm}
  \end{center}
\caption{Detail from a signal timeline 
from an observation of the standard calibrator Ceres, observed in the
C105 filter. The green line shows the 
fit to the observed signal (shown in red) from pixel 1 of the C100 
array, as found by the P32 algorithm for 
correction of the transient response behaviour of this pixel. 
5 chopper sweeps, each comprising 13 pointing directions 
(``chopper plateaus'') are shown, divided by vertical dotted lines
separated in time by a duration of 6.1s for each sweep.
The upper line in the figure shows the fine pointing flag;
whereas the first chopper sweep shown was made while the spacecraft was 
slewing between fine pointings (flag value 0), the subsequent
chopper sweeps were made on a single fine pointing (flag value 1).
The remaining horizontal lines denote masks at the full time resolution
of the data denoting glitches, readout status and whether or not a solution
for the illumination could be found by the algorithm. The overall solution
for the sky illumination for this detector pixel is given by the histogram in
black. As described in the text, this overall solution is calculated 
from a combination of individual solutions for illumination on
each successive chopper plateau, which are shown here as purple bars. 
For the very brightest sources, as shown in this example,
the derived illuminations can be up to a factor of 6 brighter than 
the raw data for pixels in the C100 array. In this example the hook 
response is well seen in each chopper sweep at the 9th and 10th chopper 
plateaus.
}
\end{figure*}

\begin{enumerate}

\item Edited Raw Data is imported, after having been corrected in PIA
for the integration ramp non-linearity.

\item The specification of chopper phase in the data is checked
and corrected where necessary.

\item The pointing is established from the instantaneous
satellite pointing information, unlike PIA which uses the commanded
pointing. This also establishes the pointing directions on the slews
between the spacecraft fine pointings. 
Internally defined ``on target flags'' are set for each spacecraft
fine pointing direction, and the central direction and sampling 
intervals directions for the P32 ``natural grid'' (see Sect.~2.1) 
are established.

\item The integration ramps are differentiated and the data is
corrected for dark current and dependence of photometry 
on reset interval, using standard calibration parameters taken
from PIA. 

\item Systematic variations of the signal according to 
position in the integration ramps are removed from the data. This 
process is called ``signal relativisation''. 
In contrast to the default pipeline or PIA analysis, this allows the use 
of all non-destructive reads, as well as providing a more complete
signal timeline for use in the correction of the transient
behaviour of the detector. The ``signal relativisation'' is 
applied such that signal variations due to the transient 
response behaviour are preserved.

\item Random noise is determined by examining the statistics of the
signals on each chopper plateau. This is necessary due to the
high readout rates of the detectors in many P32 observations. Otherwise,
the determination of noise from individual readout ramps, as done in a 
standard PIA analysis, would be insufficiently precise.

\item Spikes are detected and removed from the data in a first-stage
deglitching procedure. This operates at the full time resolution of the data
given by the non-destructive read interval.

\item A second stage deglitching procedure is performed,
operating at the time resolution of the chopper plateau 
(which comprise of at least 16 non-destructive read intervals).
This procedure (described by Peschke \& Tuffs 2002)
removes any long lived glitch ``tails''
following the spikes detected in the first-stage deglitch.

\end{enumerate}

\subsection{Transient Correction}

A data flow diagram summarising the iterative non-linear optimisation
process to solve for the sky brightnesses sampled on the P32 natural grid is
summarised in Fig.~7. 
The calculation to correct for the transient response behaviour
is made separately for each detector
pixel. The kernel of the procedure is to
solve for the illumination corresponding to the
signal on each chopper plateau. This is done in a constrained optimisation
using a binary chop scheme
for set values of the 12 detector parameters. The constraints set for
the solution in illumination define the search range for
illumination. They are set such that the solution for illumination
should be positive, and less than 10 times the value of the highest 
uncorrected signal. In practice the illuminations found on bright sources
are never more than a factor of 5 higher than the uncorrected signal,
so the determined photometry is not sensitive to the exact value of the
upper limit in the search range. 
The values for $S_{\rm 1p}$
and $S_{\rm 2p}$ needed to solve for illumination on each chopper plateau
are calculated from the response of the model to the preceding illumination
history.
\begin{figure*}[!ht]
  \begin{center}
    \epsfig{file=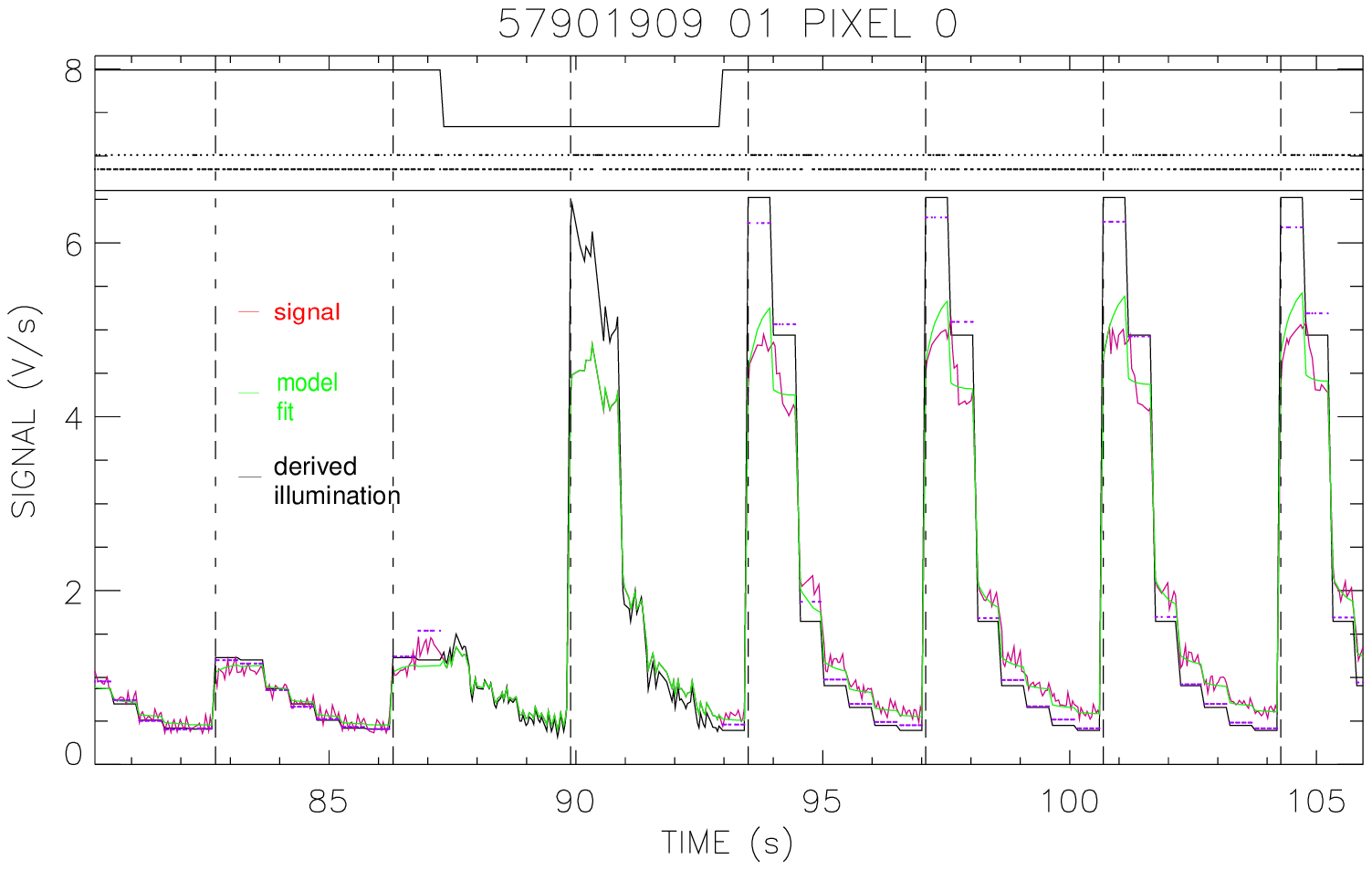, width=15.0cm}
  \end{center}
\caption{Detail from a timeline of signal versus time
from another observation of Ceres, this time observed in the
C160 filter using the C200 detector array. The key to the plot is as given 
in the caption to Fig.~8. This example shows the transition from 
a spacecraft fine pointing for which the chopper is sweeping
through beam sidelobes to a fine pointing where the chopper samples 
the beam kernel. In this case the red line corresponding to the data is
not seen during the intervening spacecraft slew, as it is
exactly overplotted by the green line showing the fit to the data
(see text). The chopper sweeps on the second fine pointing encompassing the
beam kernel show a typical enhancement of the source-background contrast 
induced by the algorithm for correction of the transient response.
}
\end{figure*}
After solving for the illumination on the current chopper
plateau (in the representation in Fig.~7, to give the {\it kth} 
value $I_K$ of the sequence of illuminations in the ``signal timeline''),
the found value $I_K$ is divided by the vignetting corresponding
to the chopper phase to obtain a new estimate of the sky brightess
for the viewed direction on the P32 natural grid. This sky brightness
estimate is then combined with any previous solutions for the 
same P32 natural grid pixel obtained with the same detector pixel from 
previous chopper plateaus, and the trial sky map 
(sampled on the P32 natural grid) is updated for this improved knowledge.
The response of the
detector pixel to the illumination history is then recalculated and
a solution is found for the illumination viewed on the next chopper 
plateau. This procedure is continued until all the signal timeline
has been processed. Then a second pass through the data
can be made (the ``second major iteration'' loop in Fig.~7), and so on, 
until the overall solution for the sky brightness distribution no 
longer changes between successive major iterations. Any sky directions
for which no solution can be found are flagged in an output mask,
sampled on the P32 natural grid.

The goodness of the solution is quantified through 
a value for $\chi^{\rm 2}$,
calculated from a comparison between the model fit and the signal timeline.
The random uncertainties assigned to each signal sample in the timeline
can optionally be recalculated from the self consistency of 
solutions for the illumination found in a given sky direction.
Examples of model fits to the signal timeline are shown in
Fig.~8 for the C100 detector and in Fig.~9 for the C200 detector.

The algorithm can also be used for datasets for which the 
readout rate was too rapid to allow all the data to be transmitted 
to the ground. In such cases the illumination history can still be
determined at all times on fine pointings, 
even where there are gaps in the signal timeline. On slews,
the illumination history during the gaps are either interpolated 
in time, or taken from the nearest previously solved direction
on the P32 natural grid. 

For most observations using the C200 detector, and observations
of moderate intensity sources made using the C100 detector, 
good fits to the entire unbroken signal timeline
can be obtained for each detector pixel. For observations of
bright sources using the C100 detector, however, 
it is often necessary to break up
the solutions into smaller segments, according to the distribution
of the sources in the field mapped. This is best achieved by 
solving separately for the portions of the signal timeline
passing through the bright sources. The most conservative
processing strategy, appropriate for observations using the C100
detector of complex brightness distributions with high source to 
background contrast, is to break up the calculation into 
time intervals corresponding to individual spacecraft fine pointings. 
In this latter case, no processing of the slews is required.

Several processing options are available for the interactive 
processing, to optimise solutions
for the particular characteristics of each dataset:

\bigskip

\noindent {\it Determination of detector starting state}

\medskip

The default operation of the program is to assume that the
detector is in equilibrium at the start of the observation,
viewing the sky direction corresponding to the first detector plateau.
This corresponds to the case $S_{\rm 1p}\,=\,S_{\rm 01}\,=\, S_{\rm \infty}$
and $S_{\rm 1p}\,=\,0$.
This will very rarely be the case, however, since an FCS measurement 
which may not be perfectly matched to the sky brightness will have been
made just seconds prior to the start of the spacecraft raster.
Therefore, the algorithm can optionally find a solution for the
detector starting state, as characterised by the values of
$S_{\rm 1p}$ and $S_{\rm 2p}$ immediately prior to the
start of the first chopper plateau. This is done by solving
for $S_{\rm 1p}$ and $S_{\rm 2p}$ for a fixed $S_{\rm \infty}$
on the first plateau, where $S_{\rm \infty}$ is determined from
the last chopper plateau in the first spacecraft pointing to view
the same sky direction. The timeline for the first spacecraft
pointing is then repeatedly processed until a stable solution for
$S_{\rm 1p}$ and $S_{\rm 2p}$ prior to the first plateau is found.
This option was found to be particularly useful for
data taken with the C200 detector. 

\bigskip

\noindent {\it ``Self calibration'' of detector parameters}

\medskip

In order to determine the primary detector parameters $\beta_{\rm 2}$
and $\tau_{\rm 2}$ in the first instance, 
``a self calibration'' processing technique was used
by which any of the 12 detector parameters (Sect.~3.1) could be determined
from observations of bright sources. The self calibration functionality
has also been implemented in the publically available P32Tools. This is
because, even though the standard parameters of 
the detector model given in Tables 1 and 2 generally give good results, 
some fine tuning of the parameters may still yield improved fits to the
signal timeline for some astronomical targets, particularly for targets with a
high contrast to the background.

Sources of any arbitrary
brightness distribution and position can be used for the self calibration. 
However, in the initial determination of the detector parameters for the 
``fast'' response of the detector, point source calibrators of
known brightness were used, and the solutions for $\tau_{\rm 2}$ and
$\beta_{\rm 2}$ were additionally constrained by the requirement that the
solutions for illumination were consistent with the known flux densities 
of the calibrators.

The self calibration procedure works by simply repeating the
optimisation process for different subsets of the 12 detector
parameters, specified by the user. Each combination
of detector parameters produces an individual solution for the sky brightness 
distribution, as well as a value for $\chi^{\rm 2}$. A search is made 
in the parameter space of the detector parameters
until a minimum is found in $\chi^{\rm 2}$. This is a lengthy
process, however, making it advisable to perform self calibrations
on limited portions of the timeline. This is often chosen to correspond
to a single spacecraft pointing where the chopper sweeps for the detector
pixel being investigated pass through bright structure. The process
can be further accelerated in cases that the repeated chopper sweeps
through a source give the same repeated signal pattern. Then, 
the optimisation can be performed on an average of the coadded chopper 
sweeps performed on a single spacecraft pointing. This 
latter technique is referred to in P32Tools as a ``composite 
self calibration''. Some examples of the useage of the self calibration
procedure are given by Schulz et al. (2002a).

\bigskip

\noindent {\it Processing of slews}

\medskip

The calculation of the illumination history for data taken
during the slews is less straightforward than for data 
taken during the spacecraft fine pointings. This is because
the instantaneous pointing directions during slews are effectively 
at random positions with respect to the P32 natural grid. Therefore,
the use of solutions for the sky brightness distribution
on the P32 natural grid found from analysis of previous fine pointings
will in general lead to inaccuracies in the determination of
the illumination history during the slews. 
There are two main approaches which can be adopted to counter this 
problem.

The first is to solve for the illumination not over a chopper
plateau, but for individual readouts in the signal timeline
during the slew. This automatically provides a solution in which
the model fit exactly matches the data, as there is only one
fitted data point for each determined illumination. An example
is given in the fit to the signal timeline for the C200 detector
shown in Fig~9. This approach is useful for observations of bright
structured sources in which the variations in illumination
with position on the slew on timescales of a chopper plateau
exceed the signal to noise on individual data samples. It is
the default processing option for the C200 detector.

In cases of low signal to noise, and for all instances where
not all data are transmitted to the ground, solving
for the average illumination on each chopper plateau is
a preferable technique. Otherwise, the values of $S_{\rm 1p}$
and $S_{\rm 2p}$ calculated just prior to the first chopper plateau
on the next fine pointing can be wildly inaccurate, which can corrupt
the solutions for the remainder of the signal timeline of the observation.
Processing on a plateau by plateau basis is the default 
processing option to calculate the illumination history during
slews for observations with the C100 detector. Another reason for
processing slews on a plateau-by-plateau basis for the C100 detector
is that the procedure to solve for illumination for individual readouts
is very susceptible to any residual glitches, and the C100
detector is more prone to glitching than the C200 detector.

\section{Photometric performance}

In this section we give examples of the photometric performance
achieved from maps made using the P32 algorithm. In all cases the
standard calibration parameters used in the signal conditioning
step prior to the correction of the transient response behaviour and
the subsequent 
\begin{figure}[htb]
\includegraphics[scale=0.6]{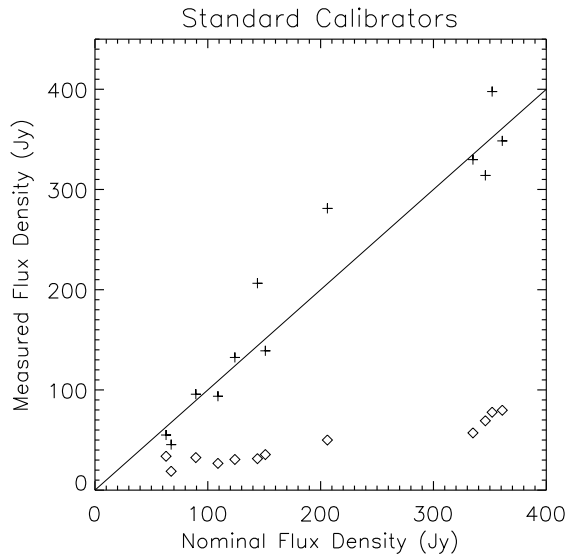}
\caption{
The measured integrated and colour-corrected flux densities 
of the standard calibrators Ceres, Vesta and Neptune
plotted against the nominal flux densities. The
observations were done in various filters using the C100 detector. The
photometry derived with and without correction for the transient
response behaviour is shown with crosses and diamonds, respectively.
The solid line shows the expected trend for equality between the
measured and nominal flux densities.
}
\end{figure}
conversion from engineering units ($V/s$) to 
astronomical units ($MJy/str$ and $Jy$) were taken from version 8.1
of PIA. To illustrate the quantitative effect of the transient correction 
we give results for datasets processed by the P32 algorithm both
with and without the transient correction. In all cases, the
corrections for the transient response behaviour were made using
the standard detector model parameters given in Tables 1 and 2.
In Sect.~5.1 we 
give examples of the effects of the transient correction on 
the derived spatially integrated flux densities. Where possible these 
integrated flux densities are also compared with flux densities
from IRAS or from source models for the standard calibrators. In Sect.~5.2
we give examples of the imaging performance.

\subsection{Integrated Flux Densities}

In general the corrections in integrated flux densities
made by the algorithm depend on the source brightness, structure, size,
the source/background ratio, and the dwell time on each chopper plateau.
The largest corrections are for bright point sources on faint backgrounds
observed with the C100 detector.

In Fig.~10 we plot the measured integrated and colour-corrected flux densities
of standard ISOPHOT calibrators (the asteroids Ceres and Vesta, and the 
planet Neptune) versus the nominal flux densities of these
sources (determined as described by Schulz et al. 2002b). Since
these same sources were used to derived the detector model, one expects
a good statistical match between the nominal and the derived flux
densities. Indeed, the integrated flux densities found after correction
for the transient behaviour of the detector lie reasonably close 
to the solid line indicating an equality between 
the measured and nominal flux densities, though with some scatter
(of around 15 percent). It can be seen that the flux densities derived from 
data uncorrected for the transient effects have been
raised by factors ranging from
$\sim\,$2 for the fainter standards to $\sim\,$5 for the brighter ones.
\begin{figure}[htb]
\includegraphics[scale=0.5]{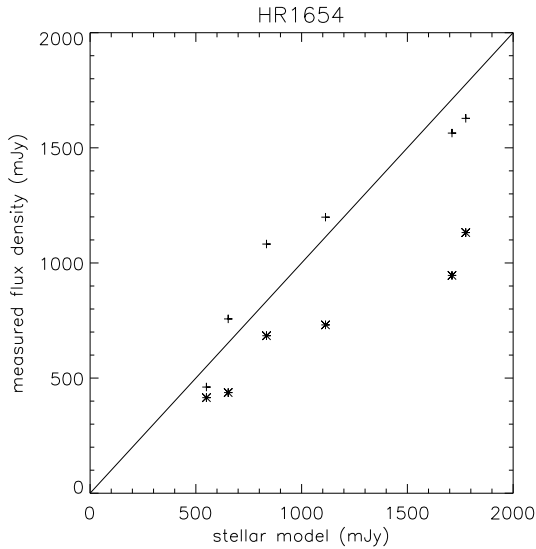}
\caption{
The measured integrated flux densities of the faint standard star HR~1654
plotted against the predicted flux densities from a stellar model. The
observations were done in various filters using the C100 detector. The
photometry derived with and without correction for the transient
response behaviour is shown with crosses and stars, respectively.
The solid line shows the expected trend for equality between the
measured flux densities and those predicted from the stellar model.
}
\end{figure}

As an example of the photometric performance for fainter point
sources we give results achieved for the faint
standard star HR~1654. This source was not used in the determination of the 
detector model parameters, so it constitutes a test of the photometric
performance of ISOPHOT in its P32 observing mode. The derived integrated flux 
densities, with and without correction for the transient response behaviour of 
the C100 detector, were compared in Fig.~11 with predicted flux densities from
a stellar model. The corrected photometry is in reasonable agreement with 
the theoretical predictions. As expected, there is again a trend for 
observations with larger detector illuminations to have larger corrections in 
integrated photometry. 

\begin{figure}[htb]
\includegraphics[scale=0.5]{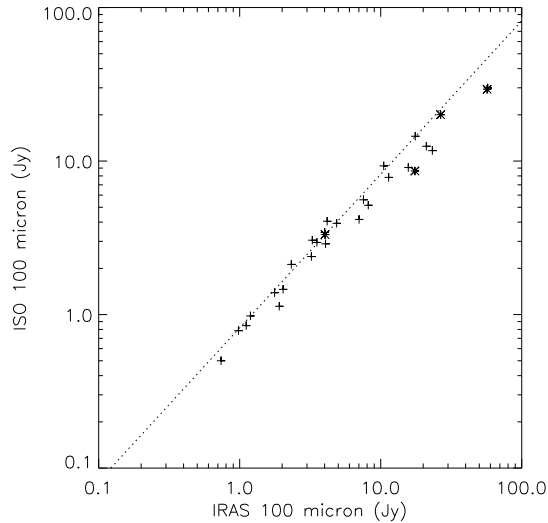}
\caption{Integrated and colour-corrected flux densities of
Virgo cluster galaxies measured in the ISO C100 filter versus
the corresponding flux densities measured by IRAS in its 
100$\,\rm \mu m$ band (taken from Fig.~7 of Tuffs et al. 2002a). The dotted
line represents the relation ISO/IRAS=0.82. The star symbols are used for
galaxies not fully covered by the spacecraft raster.}
\end{figure}

A good linear correlation is also seen between integrated flux densities
of Virgo cluster galaxies (Tuffs et al. 2002a), derived from P32 ISOPHOT 
observations processed using the P32 algorithm, and flux densities from 
the IRAS survey (Fig.~12). The interpretation of these measurements
constituted the first science application (Popescu et al. 2002) of the
P32 algorithm. The ISOPHOT observations of Virgo cluster galaxies 
were furthermore used to derive the ratios of fluxes measured by ISO to those 
measured by IRAS. The ISO/IRAS ratios were found to be
0.95 and 0.82 at 60 and 100$\,\rm \mu m$, respectively, after scaling the 
ISOPHOT measurements onto the COBE-DIRBE flux scale (Tuffs et al. 2002a).

The effect of the transient response behaviour on the integrated
flux densities of very extended sources is illustrated by the
photometry of a few bright extended nearby galaxies. 
Thus, in Table~3, Col. 5, we give the integrated flux densities 
of NGC~6946, M~51 and M~101 taken from Tuffs et al. 
(2003; in preparation). For NGC~6946 the integrated flux densities are
given here for the first time; those for M~51 and M~101 
supercede the values given by Hippelein et al. (1996). 
The standard values for the detector parameters (as given in Tables 1 and 2)
were used for the correction of the transient response.
Table~3 also
gives a comparison with the IRAS flux densities at 60 and 100$\,\rm \mu m$.
In general, the ISOPHOT photometry yields somewhat lower integrated flux
densities than the corresponding IRAS photometry, 
similar to the ISO/IRAS ratios
found for the Virgo cluster galaxies (Tuffs et al. 2002a). 
As for the point sources, we also
reduced the maps of these nearby galaxies without correcting for 
the transient response of the detectors. We were thus able to derive
the ``correction factor'' (Col. 6) indicating the ratio between the
integrated flux densities with and without correcting for the transient. 
In comparison to the point sources of similar integrated flux densities, 
these corrections factors are moderate, varying between 1.07 and 1.45.
This is because the timescale for the detector to traverse these extended
sources becomes comparable to the time constants for the transient behaviour.
The result therefore is a redistribution of flux density within the area
of the galaxies along the scan direction, which goes some way to
offset the loss of signal. We emphasise however that the remaining corrections
are systematic, rather than random effects, and that the precise value
of the correction factor for integrated flux density depends on the extent and 
morphology of the target. Furthermore, the correction factors for any
point-like sources within extended sources, such as galactic nucleii or
HII regions, will be much larger, comparable to those found for the point
source calibrators.

\begin{table*}[htb]
\caption{Photometry of selected extended sources}
\begin{tabular}{lcllccl}
\hline\hline
Source & Size    &  TDT &     Filter & Integrated      &  Correction & Integrated \\   
       & (arcmin)&      &            & Flux Density         &  Factor     & Flux Density\\
       &         &      &            &(ISOPHOT; Jy)    &             & (IRAS; Jy)$^{a}$\\
\hline
NGC6946&  9 x 7  & 05302402 &   C60  &  $111.5 \pm  6.5$  &   1.45      &     146.5\\
NGC6946&         & 05302401 &   C200 &  $365.8 \pm 11.7$  &   1.11      &        -\\
M51$^{b}$   &  9 x 6  & 17200618 &   C60  & $ 70.3 \pm  3.0$  &   1.30      &       84.9\\
M51$^{b}$   &         & 17200618 &   C100 & $162.2 \pm 12.3$  &   1.33      &      192.6\\
M51$^{b}$   &         & 17200820 &   C160 & $180.9 \pm 23.9$  &   1.07      &        -\\
M101   & 21 x 10 & 16500615 &   C60  &   $61.1 \pm  3.2$  &   1.15      &       92.9\\
M101   &         & 16500716 &   C100 &  $204.5 \pm 10.2$  &   1.12      &      263.5\\
M101   &         & 16500716 &   C160 &  $360.5 \pm 49.0$  &   1.06      &        -\\
\hline
\end{tabular}

$^{a}$  From Table 4 of Rice et al. (1988)\\
$^{b}$ Together with companion galaxy NGC 5195\\
\end{table*}

\subsection{Imaging Performance}

Fig.~13 shows an example of a brightness profile through HR~1654 at
100$\,\rm \mu m$, for data processed with and without the responsivity 
correction. Some 95$\%$ of the flux density has been recovered with the
P32 algorithm. 
Without the correction, some 30$\%$ of the integrated emission is missing
and the signal only reaches 50$\%$ of the peak illumination.
The local minimum near 60~arcsec in the Y offset is a typical hook
response artifact, where the algorithm has overshot the true solution.
This happens for rapid chopper sweeps passing through the beam kernel. 
This is a 
fundamental limitation of the detector model, which, as described in 
Sect.~3.1, does not correctly reproduce the hook response on timescales
of up to a few seconds. This problem is particularly apparent for
downwards illumination steps. The only effective antidote is to
mask the solution immediately following a transition through a bright
source peak. 
\begin{figure}[h]
  \begin{center}
    \epsfig{file=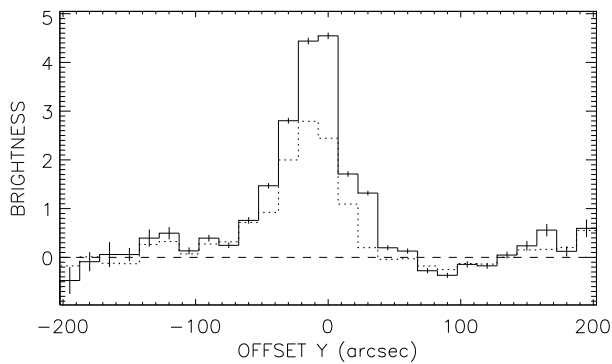, width=8.0cm}
  \end{center}
\caption{Brightness profiles (in MJy/str) along the Y spacecraft direction 
through the standard star HR~1654
at 100$\,\rm \mu m$. The solid line represents the brightness profile
obtained after processing with the P32 algorithm, while the dotted line 
shows the profile derived from identically 
processed data, but without correction for the transient response of
the detector.
}
\end{figure}

Another effect of the inability to model the
hook response is that the beam profile becomes somewhat distorted.
This also has the consequence, that for observations of bright sources,
the measured FWHM can become narrower than for the true point spread function,
as predicted from the telescope optics and pixel footprint. 
We emphasise that, because of the beam distortion, the flux density of
bright point sources cannot be found from the peak surface brightness,
but instead should be determined by integrating the map. An example
of an extremely bright point source reduced using the P32 algorithm is 
given in Fig.~14, depicting maps of Ceres in the C105 
filter, made with and without the correction for the transient 
response of the C100 detector.  

\begin{figure}[htb]
\includegraphics[scale=0.72]{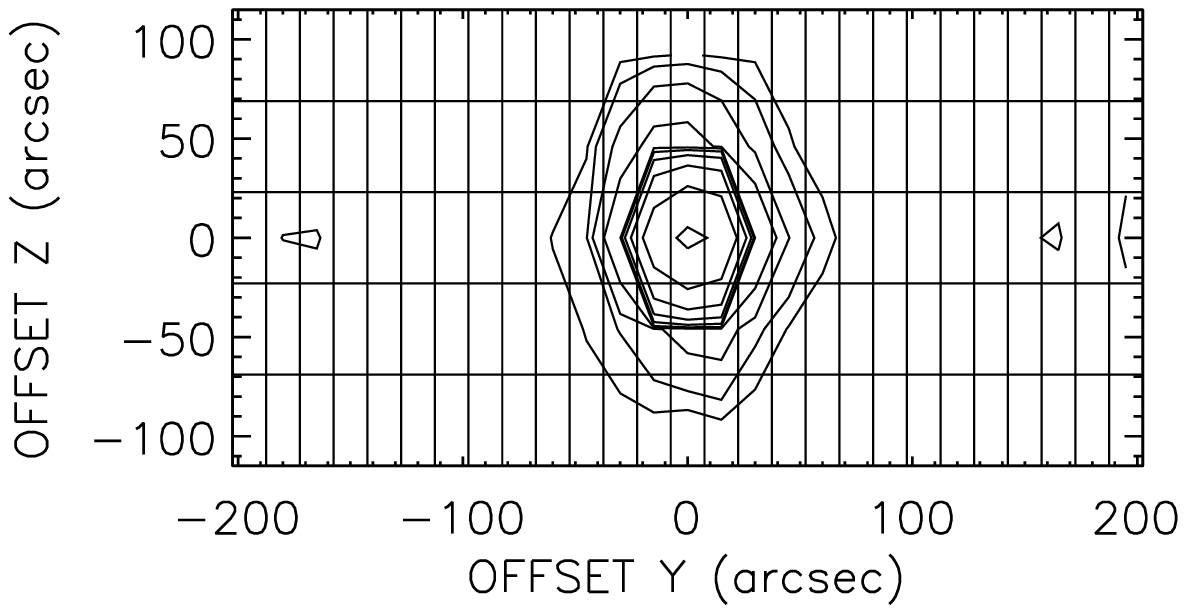}
\includegraphics[scale=0.72]{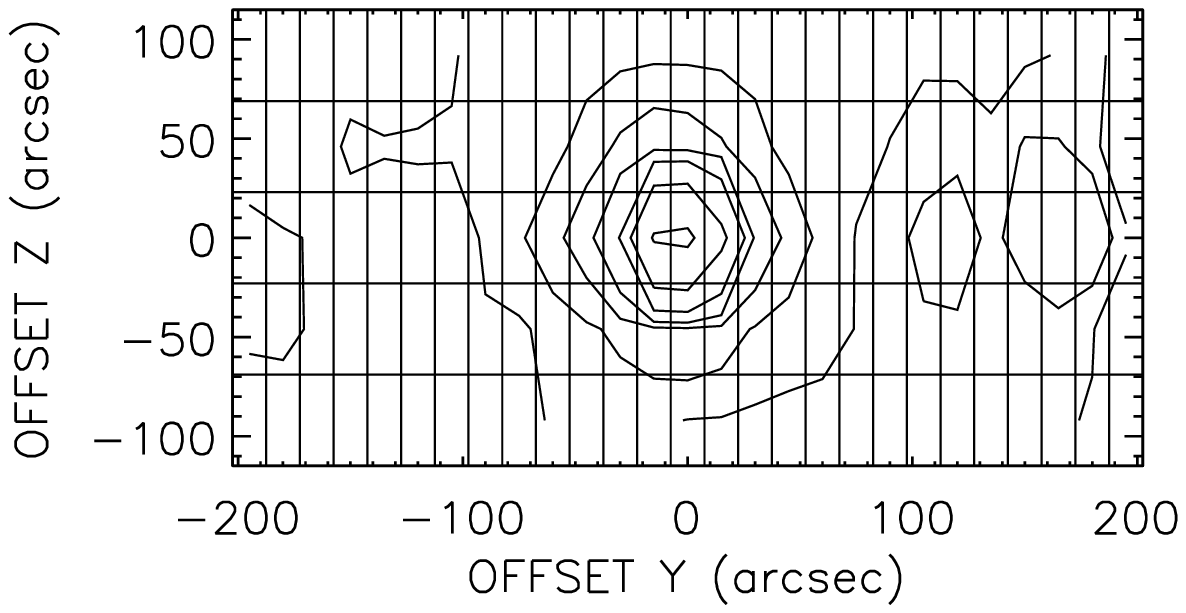}
\caption{Top: Contour map of Ceres in the C105 filter after
responsivity drift correction. The contour levels are $5\,.\,2^{n}$\,MJy/sr
for all integers n with $1\,\le\,n\,\le\,10$. The
measured integrated flux density
after background subtraction is $98.4$\,Jy\,$\pm\,0.2$\,Jy (random) 
$\,\pm\,6\%$ (systematic). The actual flux density of this standard
calibrator (from a stellar model) is 109\,Jy. As in Fig.~2, the
the map pixels of the P32 natural grid have been overlaid.
Bottom: Contour map of Ceres in the C105 filter without
correction for the transient response behaviour of the detector.
The contour levels are $5\,.\,2^{n}$\,MJy/sr
for all integers n with $1\,\le\,n\,\le\,7$. The 
The measured integrated flux density
after background subtraction is $28.0$\,Jy\,$\pm\,0.07$\,Jy (random) 
$\,\pm\,4\%$ (systematic)}
\end{figure}
\begin{figure}[htb]
\includegraphics[scale=0.72]{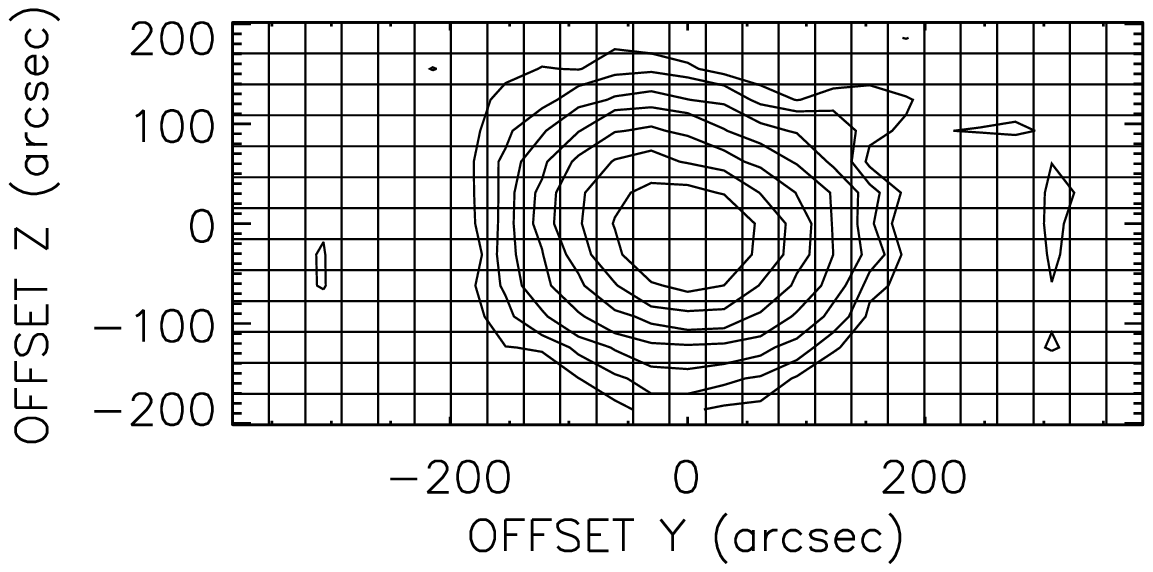}
\includegraphics[scale=0.72]{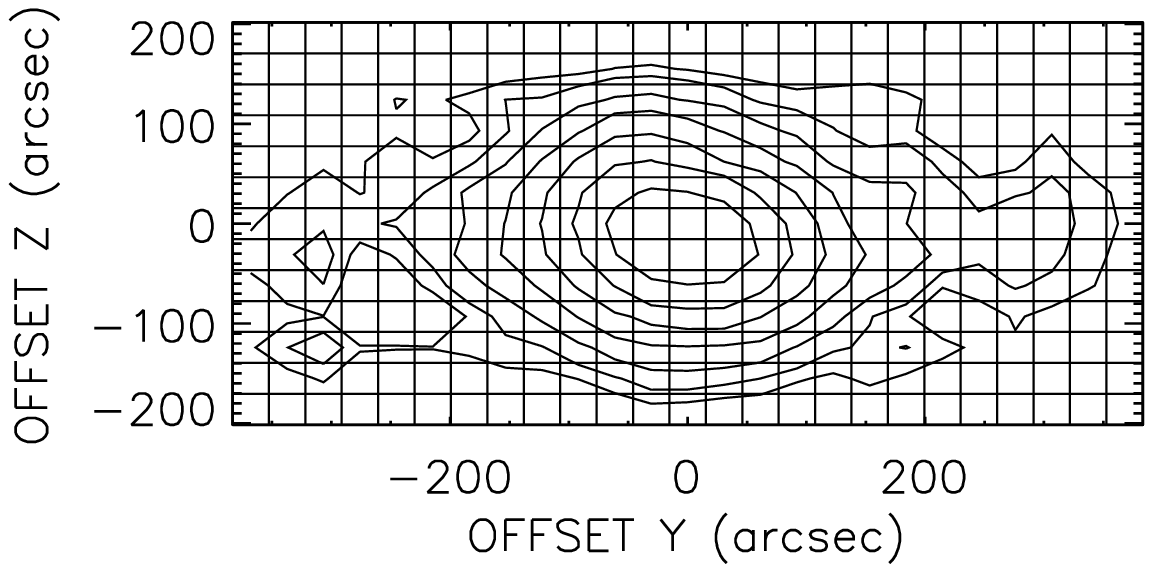}
\caption{The interacting galaxy pair KPG~347 observed in the
C160 filter. Top: after processing with the P32 algorithm. 
Bottom: with identical
processing, except that the correction for the transient 
response behaviour of the detector has been omitted.
The map pixels corresponding to the P32 natural grid (see Sect.~2.1)
have been overlaid. In both maps, contours are logarithmic, at levels 
1, 2, 4, 8, 16, 32, 64~MJy/sr. The peak brightnesses
are 126 and 109~MJy/str for the maps with and without
processing with the P32 algorithm, respectively.
}
\end{figure}

\begin{figure}[htb]
\includegraphics[scale=0.65]{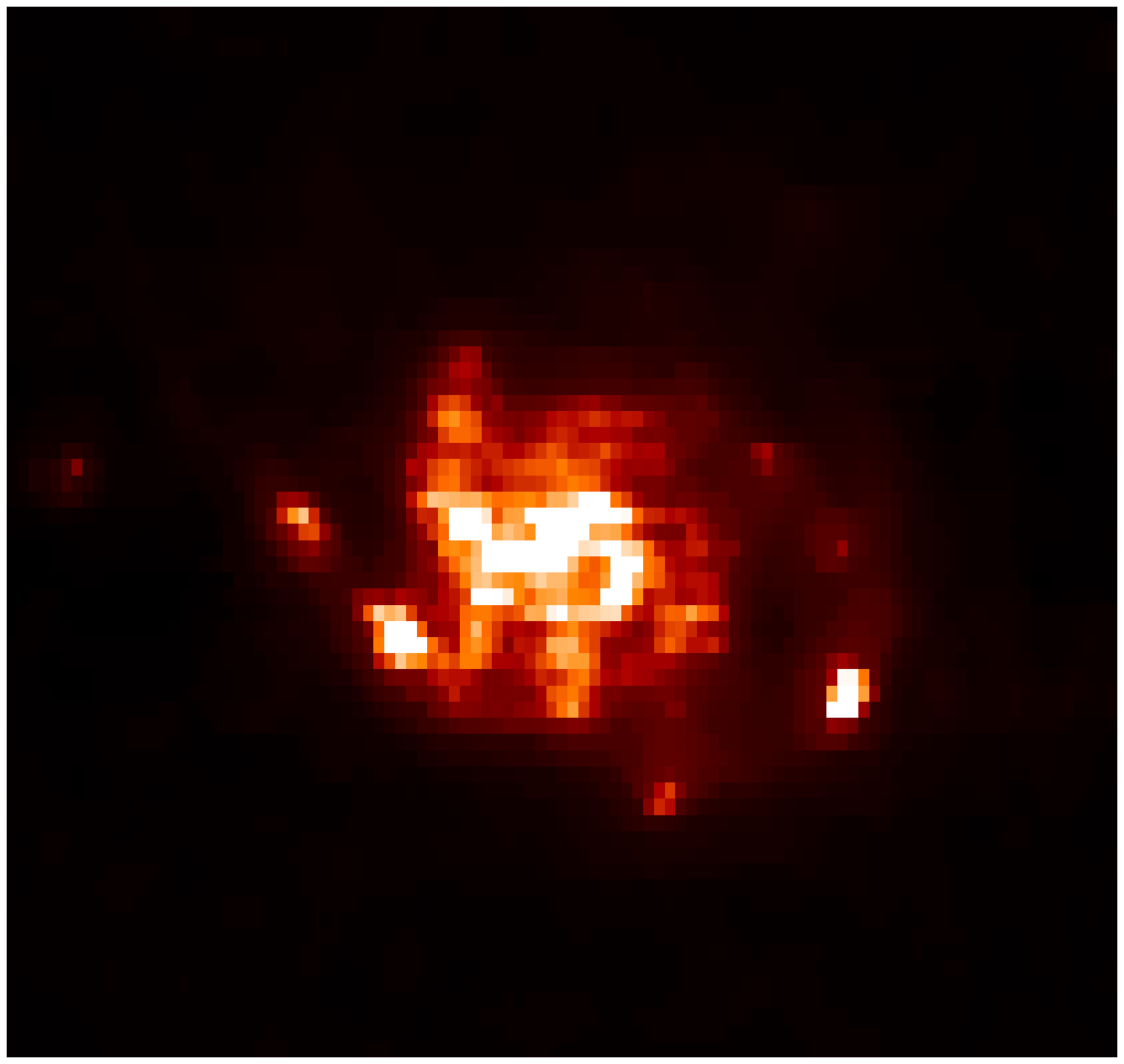}
\caption{A $27\times27$ arcmin field containing the galaxy M101,
as mapped in the C100 filter, after processing with the P32 algorithm. 
The map sampling is $15\times23$ arcsec.
 }
\end{figure}

Despite the limitations due to the lack of a proper modelling of the
hook response, the algorithm can effectively correct for artifacts 
associated with the transient response on timescales from a few seconds
to a few minutes. This is illustrated in Fig.~15 
by the maps of the interacting
galaxy pair KPG~347 in the C200 filter 
(again, after and before correction for the transient response
of the detector, respectively). 
The uncorrected map shows a spurious elongation in
the direction of the spacecraft Y coordinate,
which is almost completely absent in the corrected map. If uncorrected,
such artifacts could lead to false conclusions about the brightness
of FIR emission in the outer regions of resolved sources. Also visible 
are artifacts at 
Y$\,=\,\pm\,$310~arcsec. Similar artifacts are also seen at the same offset 
in other P32 observations made with the C200 detector, with position and
amplitude relative to the central source independent of the
filter used. These features are
attributable to optical effects peculiar to the P32 mode, in which a small 
fraction of the light from a centrally positioned compact source is 
deflected away from the main beam into the vignetted part of the field
of view for a certain combination of the offset in spacecraft fine pointing 
and chopper phase. 
Although these artifacts have nothing to
do with the transient response behaviour of the detector, they are alleviated
by the application of the P32 algorithm when solutions for detector 
illumination are converted into sky brightnesses, due to the
vignetting correction made in this processing step.

The image of M~101 in Fig.~16, made using the C100 detector with
the C100 filter, is given as a state of the art 
example of what can be achieved with a careful interactive 
processing of P32 data using the P32 algorithm. In addition to the
transient response corrections and a masking of residual
hook response artifacts, a time dependent flat field 
has been applied. The spiral structure of the galaxy, with
embedded HII region complexes and a component of diffuse
interarm emission can clearly be seen.

\begin{acknowledgements}

This work was supported by grant 50-QI-9201 of the Deutsches Zentrum f\"ur
Luft- und Raumfahrt. We would like to acknowledge all those who have helped 
us in many ways in the development of the algorithm described here. 
Richard Tuffs would like to thank his colleagues at the 
Max-Planck-Institut f\"ur Kernphysik, in particular 
Prof. Heinrich V\"olk, for their support and encouragement. Dr. Cristina
Popescu is thanked for the careful reading of the manuscript 
and comments. Thanks are
also due to Prof. Rolf Chini for his input to the definition of the
original concept of the P32 AOT. 
We have also benefited from many useful discussions with 
Drs. R. Laureijs, S. Peschke and B. Schulz and the team in the 
ISO data centre at Villafranca, with Prof. D. Lemke, Drs. U. Klaas, M. Haas and
M. Stickel at the ISOPHOT data centre at the Max-Planck-Institut f\"ur 
Astronomie, and with Drs. N. Lu and I. Khan at the Infrared Processing
and Analysis Center. We thank the referee for his helpful and 
astute comments.

\end{acknowledgements}

\end{document}